\documentclass[preprint,showpacs,amsmath,amssymb,prd,nofootinbib]{revtex4}
\usepackage{epsf}
\usepackage{dcolumn}
\usepackage{amsmath}
\usepackage{amssymb}
\usepackage{graphicx}

\usepackage[breaklinks=true,colorlinks=true]{hyperref}

\begin{document}

\title{Longitudinal electromagnetic waves in the framework of standard classical electrodynamics}

\author{Volodimir Simulik\footnote{Email: vsimulik@gmail.com}}

\affiliation{Institute of Electron Physics of Ukrainian National Academy of Scientists,
Uzhgorod 88017, Ukraine}


\begin{abstract}
Theoretical comment for the registration of longitudinal electric waves in interacting laser beams is given. Recent information on longitudinal electric and scalar waves in plasma, plasmons, waveguides, antennas and nano-structures is considered. The link between the longitudinal electromagnetic waves and the system of Maxwell equations is demonstrated. The longitudinal wave component of the electric field strength vector is found as the exact solution of the standard Maxwell equations with specific gradient-type case of electric current and charge densities. The corresponded scalar wave is found as well. The necessary comment on the paper of K.J. van Vlaenderen and A. Waser is given. The possible experimental situation is discussed briefly. Following the Bialynicki-Birula's proposition to give the name for the well known complex 3-vector of electromagnetic field strengths, the suggestion to call the much more interesting 4-vector of this kind as Riemann--Silberstein--Darwin vector is given. 
\end{abstract}

\pacs {11.10.-z, 41.20.-q, 41.20.Jb}

\maketitle


\section{Introduction}

The theoretical and experimental description of the longitudinal electromagnetic waves is a subject of the different level discussions. There is a lot of indications of the existence of such waves in different media. Note, e.g., the longitudinal waves in plasmons and in plasma. Nevertheless, these oscillations only sometimes are considered as the electromagnetic ones. The longitudinal electromagnetic waves are well known to waveguide engineers. Let us mark especially the recent observations of the longitudinal electric wave in an interacting laser beams. On the other hand, it is well known fact that the Maxwell equations for the free electromagnetic field contain only the transverse solutions.

Recently in \cite{Miyaji} a simple experimental method for generating an intense longitudinal electric field from transverse electromagnetic waves (laser pulses) with radially symmetric polarization has been presented. The laser-generated longitudinal electric field was observed in two dimensions and distinguished from the transverse component using the optical Kerr shutter method. Authors of \cite{Dorn} experimentally demonstrated that a radially polarized field can be focused to a spot size significantly smaller than for linear polarization. For strong focusing, a radially polarized field leads to a longitudinal electric field component at the focus which is sharp and centered at the optical axis. Experimental and theoretical investigations in \cite{Niziev} demonstrates that the longitudinal field component has the maximal amplitude where the "`usual"' field component is zero. The high-power industrial CO$_{2}$ laser is applied. Authors of \cite{Tidwell} studied that, when focused, a radially polarized beam has a net longitudinal field useful for particle acceleration and, perhaps, other unique applications.

It is well known that the electromagnetic field in the waveguide is not purely transverse, but has longitudinal components. Moreover, on the basis of waveguides "Longitudinal-scalar electromagnetic wave radiating device RU 2287212 C1" has been suggested \cite{Kuzn}.

The author's of theoretical article \cite{Petrov} have studied the problem of plasmons in a QED vacuum. It has been shown that the bosonized version of (1 + 1)-dimensional QED admits the existence of classical stable time-periodic solutions, i.e., standing waves of the longitudinal electric field and the vacuum’s polarization density. Note that assertions about longitudinal electromagnetic waves in plasma are known from  \cite{Kovr, Bogdan}.

In \cite{Pekar} the theory of electromagnetic waves in a crystal, where excitons are produced, have been considered. Strictly-longitudinal electric waves are shown to exist in such crystal. Recently in \cite{Dats} the properties of longitudinal electromagnetic oscillations in metals and their excitation at planar and spherical surfaces have been studied.

The radiation of longitudinal electric waves by antennas was studied recently in theoretical article \cite{Umul}. More exactly, the radiation of electromagnetic fields by an electric charge density, distributed uniformly on the surface of a conducting sphere, was the subject of investigation. The resultant electromagnetic field has longitudinal and transverse components. An important result is that the curl of the magnetic field gives only the transverse electric field component that has 1/r dependence. Also the longitudinal electric field has a spatial dependence of 1/r and does not interact with the transverse magnetic field, since it is curl-free.

In \cite{Monstein} scalar potential wave together with a longitudinal electric field $\overrightarrow{E}$ in the direction of propagation has been detected by a ball antenna. Nevertheless, the authors of \cite{Monstein,Bray,Rebilas,Butterworth} have been doubted in this result both from the theoretical \cite{Bray, Rebilas} and experimental \cite{Butterworth} points of view. The authors of \cite{Butterworth} tried to repeat the experiment \cite{Monstein} without any success. In \cite{Rebilas} it is proved that the authors of \cite{Monstein} detected classical TEM waves emitted by currents flowing in the Earth and launched by the ball antenna used in the experiment. 

The goal of this paper is to contribute in the problem by means of standard classical electrodynamics, e.g., by the exact solutions of well defined physical and mathematical systems of the Maxwell equations. Below we presented the result that the longitudinal wave component of the electric field strength vector (together with longitudinal scalar wave) is the exact solution of the standard Maxwell equations with specific gradient-type case of electric current and charge densities. The given solution can be verified by anyone after individual substitution into standard Maxwell system with corresponding sources. We explain below why this solution was not noticed before. Our results known from the year 1997 are also applied.

Note that in literature one can find some confusion in using the notions "longitudinal electric wave" and "longitudinal wave component of the electric field strength vector". It leads to the confusion in understanding.

The question about longitudinal electromagnetic waves is the old problem of classical electrodynamics. There is a large probability that N. Tesla was the first in such investigations. Nevertheless, today it is difficult to distinguish the legends about N. Tesla from his real investigations, see, e.g., \cite{Nedic}. The first known theoretical model has been considered in \cite{1941}. Longitudinal electromagnetic waves between parallel plates were investigated.

The hypothesis about the longitudinal electromagnetic waves, which presence in the mathematical formalism follows from the massless Dirac equation (on the basis of a link between the massless Dirac and slightly generalized Maxwell equations), was suggested in \cite{Hvorost}. The start in \cite{Hvorost} was related to the results \cite{Simulik-1}. Following \cite{Hvorost,Simulik-1}, in \cite{Kr-Sim-1,Krivsky-Sim-2} we presented our preliminary point of view on this problem. The general solution of such kind of the Maxwell equations has been found. The conclusion that the longitudinal electric waves can exist within the framework of standard Maxwell electrodynamics in the case, when the electric currents and charges have specific gradient-like form, has been suggested. Such longitudinal electric wave appears \cite{Kr-Sim-1,Krivsky-Sim-2} in the pair with scalar wave.

However, the problem on the longitudinal electromagnetic waves can be considered independently (without any relation to the massless Dirac equation and the Maxwell equations in the Dirac-like form). Indeed, in \cite{Kr-Sim-1,Krivsky-Sim-2} we have solved the Maxwell equations directly by the Fourier method. This result of \cite{Kr-Sim-1,Krivsky-Sim-2} was considered briefly in \cite{Simulik-2} as well. Below the longitudinal wave component of the electric field strength vector $\overrightarrow{E}$ is found as the exact solution of the standard Maxwell equations with specific partial case of electric current and charge densities. The found solution contains also the longitudinal scalar wave. The necessary physical interpretation and suggestions for observation are given. 

In the period after our publications \cite{Kr-Sim-1,Krivsky-Sim-2} an interest to the problem of longitudinal electromagnetic waves in electrodynamics has been arose. Other approaches to the problem should be briefly mentioned as well, see, e.g., \cite{Dubna,Hadronic} and brief review of the publications \cite{Miyaji,Dorn,Niziev,Tidwell,Kuzn,Petrov, Dats,Umul,Monstein,Bray,Rebilas,Butterworth,Nedic,1941} given here in the text above.

The theoretical approach of \cite{Hadronic} should be considered especially. The "generalization of classical electrodynamics to admit a scalar field and longitudinal waves" of \cite{Hadronic} was published before in our papers \cite{Kr-Sim-1,Krivsky-Sim-2}. However, the article \cite{Hadronic} does not contain the references on \cite{Kr-Sim-1,Krivsky-Sim-2}. The necessary comment for this "result" of \cite{Hadronic} is given below in section 4 (see also the report at the international conference \cite{Simulik-Gor-Z}).

Furthermore, the hasty interpretation given in \cite{Hadronic} about the generalization of electrodynamics lead to the appearance of a number of followers, which continue the "generalization" and "improvement" of the Maxwell electrodynamics. Note that one of the goals of this paper is to explain the results of \cite{Kr-Sim-1,Krivsky-Sim-2} in details and to demonstrate that the description of longitudinal electric and scalar waves and generalization of the Maxwell electrodynamics are the different problems. The evident conclusion is that the standard classical electrodynamics is working well in the region of classical phenomena in general and in description of the longitudinal electric (and scalar) waves in particular.    

Here below the contemporary status of our original approach started from \cite{Simulik-1,Kr-Sim-1,Krivsky-Sim-2} is presented. 

\section{Exact solution of the Maxwell system with gradient-type electric sources}\label{Exact}

As ordinarily in field theory, for the dielectric constant $\varepsilon$, magnetic permeability $\mu$, Planck constant and velocity of light the system of units $\varepsilon=\mu=\hbar=c=1$ is chosen. Further, in the Minkowski space-time $\mathrm{M}(1,3)=\{x\equiv(x^{\mu})=(x^{0}=t, \, \overrightarrow{x}\equiv(x^{j}))\}; \quad \mu=\overline{0,3}, \, j=1,2,3,$ the variables $x^{\mu}$ denote the Cartesian (contra-variant) coordinates of the points of the physical space-time in the arbitrary-fixed inertial reference frame. The metric tensor in Minkowski space-time $\mathrm{M}(1,3)$ is given by $g^{\mu\nu}=g_{\mu\nu}=g^{\mu}_{\nu}, \, \left(g^{\mu}_{\nu}\right)=\mathrm{diag}\left(1,-1,-1,-1\right); \quad x_{\mu}=g_{\mu\nu}x^{\mu},$ and summation over the twice repeated indices is implied.

In our main calculations we use the rigged Hilbert space $\mathrm{S}^{3,4}\subset\mathrm{H}^{3,4}\subset\mathrm{S}^{3,4*}$, where the Schwartz test function space $\mathrm{S}^{3,4}$ is dense in the Schwartz generalized function space $\mathrm{S}^{3,4*}$  and $\mathrm{H}^{3,4}$ is the quantum-mechanical Hilbert space of 4-component functions over $\mathrm{R}^{3}\subset \mathrm{M}(1,3)$.

Consider the Maxwell equations in the form

\begin{equation}
\label{Main}
\partial_{0}\overrightarrow{E}-\mathrm{curl}\overrightarrow{H}=-\mathrm{grad}E^{0}, \quad \partial_{0}\overrightarrow{H}+\mathrm{curl}\overrightarrow{E}=0,
\end{equation}
$$ \mathrm{div}\overrightarrow{E}=-\partial_{0}E^{0}, \quad \mathrm{div}\overrightarrow{H}=0.$$
The first description and application of the form (\ref{Main}) was given in \cite{Kr-Sim-1,Krivsky-Sim-2}. Compare with the standard Maxwell equations for the free field
\begin{equation}
\label{Free}
\partial_{0}\overrightarrow{E}-\mathrm{curl}\overrightarrow{H}=0, \quad \partial_{0}\overrightarrow{H}+\mathrm{curl}\overrightarrow{E}=0, \quad \mathrm{div}\overrightarrow{E}=0, \quad \mathrm{div}\overrightarrow{H}=0,
\end{equation}
and with the standard Maxwell system with electric sources. 

From the physical point of view, equations (\ref{Main}) are the partial case of the standard Maxwell equations (the partial case of a standard classical electrodynamics). In this case, the current $\overrightarrow{j}(x)$ and charge $\rho(x)$ densities have the form  

\begin{equation}
\label{Sources}
\overrightarrow{j}(x)=-\mathrm{grad}E^{0}(x), \quad \rho(x)=-\partial_{0}E^{0}(x).
\end{equation}

The mathematical point of view shows that objects (\ref{Main}) and (\ref{Free}) are the different systems of partial differential equations. Indeed, the general solution of the system (\ref{Free}) is well known and is given by the transverse electromagnetic waves only. The presence of the longitudinal solutions \cite{Kr-Sim-1,Krivsky-Sim-2} of the system (\ref{Main}) follows from the existence of additional partial derivatives $\partial_{0}E^{0}$ and $\mathrm{grad}E^{0}$ in (\ref{Main}). Therefore, any contradiction with mathematical physics is absent. On the other hand the restriction on the form (\ref{Sources}) of sources is absent in classical electrodynamics. Thus, the contradiction with classical electrodynamics is absent as well. It is the reason why the result of \cite{Kr-Sim-1,Krivsky-Sim-2, Simulik-Gor-Z} on the presence of longitudinal electric and scalar waves is quite possible in the framework of ordinary classical electrodynamics.

Consider at first the general solutions of the system (\ref{Free}):
\begin{equation}
\label{Sol2}\overrightarrow{E}(x)=\frac{1}{(2\pi)^{\frac{3}{2}}}\int d^{3}k\sqrt{\frac{\omega}{2}}\left\{\left[c^{1}_{\vec{k}}\vec{e}_{1}+c^{2}_{\vec{k}}\vec{e}_{2}\right]e^{-ikx}+\left[c^{*1}_{\vec{k}}\vec{e}_{1}^{\,*}+c^{*2}_{\vec{k}}\vec{e}_{2}^{\,*}\right]e^{ikx}\right\},
\end{equation}
$$\overrightarrow{H}(x)=\frac{i}{(2\pi)^{\frac{3}{2}}}\int d^{3}k\sqrt{\frac{\omega}{2}}\left\{\left[c^{1}_{\vec{k}}\vec{e}_{1}-c^{2}_{\vec{k}}\vec{e}_{2}\right]e^{-ikx}-\left[c^{*1}_{\vec{k}}\vec{e}_{1}^{\,*}-c^{*2}_{\vec{k}}\vec{e}_{2}^{\,*}\right]e^{ikx}\right\}.$$
Here $(\overrightarrow{E}(x), \, \overrightarrow{H}(x))$ are the real electric and magnetic field strengths, $c^{1}_{\vec{k}}, \, c^{2}_{\vec{k}}$ are the complex quantum-mechanical momentum-helicity amplitudes of a photon (the amplitudes of the transverse electromagnetic waves),
\begin{equation}
\label{Not5} kx \equiv \omega t-\vec{k}\vec{x}, \quad \omega \equiv \sqrt{\vec{k}^{2}},
\end{equation}
and the 3-component basis vectors $(\vec{e}_{1}, \, \vec{e}_{2}, \, \vec{e}_{3})$, which, without any loss of generality, can be taken as
\begin{equation}
\label{Not6} \vec{e}_{1}=\frac{1}{\omega \sqrt{2(k^{1}k^{1}+k^{2}k^{2})}}\left|
{{\begin{array}{*{20}c}
 \omega k^{2}-ik^{1}k^{3} \hfill  \\
 -\omega k^{1}-ik^{2}k^{3} \hfill  \\
 i(k^{1}k^{1}+k^{2}k^{2}) \hfill  \\
 \end{array} }} \right|, \quad \vec{e}_{2}=\vec{e}^{\,*}_{1}, \quad \vec{e}_{3}=\frac{\vec{k}}{\omega},
\end{equation}
are the eigen vectors of the quantum-mechanical helicity operator for the spin s = 1.

The general solution (\ref{Sol2}) is ordinarily found by the Fourier method, and the normalization factor $C \equiv \sqrt{\frac{\omega}{2(2\pi)^{3}}}$ in (\ref{Sol2}) is taken from the condition   
\begin{equation}
\label{Energy-1} P_{0}=\frac{1}{2}\int d^{3}x\left(\overrightarrow{E}^{2}+\overrightarrow{H}^{2}\right)=\int d^{3}k \, \omega \left(\left|c^{1}_{\vec{k}}\right|^{2}+\left|c^{2}_{\vec{k}}\right|^{2}\right),
\end{equation}
\noindent i.e., from the demand to have the dimension of energy. The conservation law of momentum is given by the Pointing vector
\begin{equation}
\label{Pointing-1}
\overrightarrow{P}=\int d^3x(\overrightarrow{E}\times \overrightarrow{H}).
\end{equation}

It is interesting to find the general solution of the system (\ref{Main}), which is expected to have another (maybe not only transverse) form. The Fourier method in the corresponding rigged Hilbert space leads to the general solution (some details on the useful functional spaces are given in \cite{Simulik-Nova}):

\begin{small}
$$\overrightarrow{E}(x)=\frac{1}{(2\pi)^{\frac{3}{2}}}\int d^{3}k\sqrt{\frac{\omega}{2}}\left\{\left[c^{1}_{\vec{k}}\vec{e}_{1}+c^{2}_{\vec{k}}\vec{e}_{2}+\alpha_{\vec{k}}\vec{e}_{3}\right]e^{-ikx}+\left[c^{*1}_{\vec{k}}\vec{e}_{1}^{\,*}+c^{*2}_{\vec{k}}\vec{e}_{2}^{\,*}+\alpha^{*}_{\vec{k}}\vec{e}_{3}^{\,*}\right]e^{ikx}\right\},$$
\end{small}
\begin{equation}
\label{Sol-1}\overrightarrow{H}(x)=\frac{i}{(2\pi)^{\frac{3}{2}}}\int d^{3}k\sqrt{\frac{\omega}{2}}\left\{\left[c^{1}_{\vec{k}}\vec{e}_{1}-c^{2}_{\vec{k}}\vec{e}_{2}\right]e^{-ikx}-\left[c^{*1}_{\vec{k}}\vec{e}_{1}^{\,*}-c^{*2}_{\vec{k}}\vec{e}_{2}^{\,*}\right]e^{ikx}\right\},
\end{equation}
$$E^{0}(x)=\frac{1}{(2\pi)^{\frac{3}{2}}}\int d^{3}k\sqrt{\frac{\omega}{2}}\left(\alpha_{\vec{k}}e^{-ikx}+\alpha^{*}_{\vec{k}}e^{ikx}\right).$$

\noindent It is easy to see that here the electric field strength $\overrightarrow{E}(x)$ contains (together with the ordinary transverse waves) the longitudinal wave as well. This longitudinal electric wave is determined by the amplitude $\alpha_{\vec{k}}$. The scalar function $E^{0}(x)$ specifies the electric current and charge densities in the Maxwell system (\ref{Main}). The scalar wave $E^{0}(x)$ is longitudinal as well and is determined by the amplitude $\alpha_{\vec{k}}$.

The conservation laws of energy and momentum have the forms
\begin{equation}
\label{Energy-2}
P^0=\frac 12\int d^3x(\overrightarrow{E}^2+\overrightarrow{H}^2+E_0^2), \quad \overrightarrow{P}=\int d^3x(\overrightarrow{E}\times \overrightarrow{H}-\overrightarrow{E}E^0).
\end{equation}

\noindent For the system of electromagnetic and scalar field $(\overrightarrow{E}, \overrightarrow{H}, E^0)$ the forms (\ref{Energy-2}) of conservation laws are evident. Note that in another possible interpretation in the framework of pure electrodynamics the expressions (\ref{Energy-2}) are only the theoretical hypotheses, which needs the experimental confirmation. 

The validity of the solution (\ref{Sol-1}) can be verified by the direct substitution of (\ref{Sol-1}) into equations (\ref{Main}).

\section{Exact solution of the Maxwell system with gradient-type electric and magnetic sources}\label{Exact-2}

Consider the Maxwell-like system of equations
\begin{equation}
\label{General}
\partial_{0}\overrightarrow{E}-\mathrm{curl}\overrightarrow{H}=-\mathrm{grad}E^{0}, \quad \partial_{0}\overrightarrow{H}+\mathrm{curl}\overrightarrow{E}=-\mathrm{grad}H^{0},
\end{equation}
$$ \mathrm{div}\overrightarrow{E}=-\partial_{0}E^{0}, \quad \mathrm{div}\overrightarrow{H}=-\partial_{0}H^{0}.$$
\noindent Equations (\ref{General}) contain both electric and magnetic gradient-type current and charge densities. System (\ref{General}) is directly related to the massless Dirac equation (see e.g. the consideration in \cite{Simulik-1, Simulik-Nova,Simulik-Natur,Broglie,Poland} and the references therein). Note that the first consideration of the system (\ref{General}) and its relationship to the Dirac equation with $m=0$ was given by C.G. Darwin \cite{Darwin} in the year of 1928.

Our interest \cite{Simulik-Nova, Broglie,Poland} to the equations (\ref{General}) follows from the results that (\ref{General}) is the most symmetrical system among the different possible forms of the Maxwell equations. Note that J.C. Maxwell derived a system of equations for describing electromagnetic phenomena on the basis of a generalized rewriting of all known electrodynamics laws of Faraday, Ampere, Weber, etc., as well as from the principle of symmetry. Looking for the equations for inneratomic problems in the framework of classical electrodynamics we have recalled Maxwell's idea about symmetry principle. By analogy with Maxwell's suggestions in the articles \cite{Broglie,Poland}, perhaps the most symmetrical form of the Maxwell equations with 256 dimensional invariance algebra is considered. It is precisely this system of Maxwell equations that is directly related to the massless Dirac equation and can describe the spectrum of hydrogen. The massless Dirac equation follows from such non-ordinary Maxwell equations. Therefore, we test the form of the Maxwell equations (\ref{General}) on the subject of the description of the longitudinal electromagnetic waves.

In the terms of 3-component complex function $\overrightarrow{\mathcal{E}}= \overrightarrow{E} -i\overrightarrow{H}$ (Riemann--Silberstein vector \cite{Silberstein,Silberstein-2} the system of equations (\ref{General}) can be rewritten in the form
\begin{equation}
\label{Complex-1}
\partial_{0}\overrightarrow{\mathcal{E}}-i\mathrm{curl}\overrightarrow{\mathcal{E}}=\overrightarrow{j}, \quad \mathrm{div}\overrightarrow{\mathcal{E}}=\rho.
\end{equation}
Notation $\overrightarrow{\mathcal{E}}= \overrightarrow{E} -i\overrightarrow{H}$ was applied to electrodynamics in \cite{Silberstein}. The Riemann contribution can be found in \cite{Silberstein-2}. The first application to the photon wave function was suggested by E. Majorana, see \cite{Majorana}. In our papers on electrodynamics \cite{Simulik-1,Kr-Sim-1,Krivsky-Sim-2,Simulik-2, Simulik-Gor-Z, Simulik-Natur,Broglie,Poland} the vector $\overrightarrow{\mathcal{E}}$, as the $(\overrightarrow{E}, \overrightarrow{H})$ pair, is called the electromagnetic field.

We consider the special partial case, when in equations (\ref{Complex-1}) all the components $j^{\mu}$ of the 4-current density
\begin{equation}
\label{Lab13}
j\equiv j^{\mu}=(\rho, \overrightarrow{j}): \rho=j^{0}, \overrightarrow{j}=(j^{\ell}),
\end{equation}
\noindent are defined by one scalar function $\varphi$ according to the formula
\begin{equation}
\label{Lab14}
j^{\mu}=-\partial_{\mu}\varphi: \rho=-\partial_{0}\varphi, \overrightarrow{j}=-\mathrm{grad}\varphi.
\end{equation}
\noindent The current density (\ref{Lab14}) is called the gradient-like one (or the gradient-like source).

In the terms of 4-component function
\begin{equation}
\label{Lab15}
\mathcal{E}\equiv (\mathcal{E}^{\mu}): \mathcal{E}^{0}=E^{0}-iH^{0}\equiv\varphi, \, (\mathcal{E}^{\ell})\equiv \overrightarrow{\mathcal{E}}= \overrightarrow{E} -i\overrightarrow{H}, 
\end{equation}
\noindent equations (\ref{Complex-1}) with gradient-like sources (\ref{Lab14}) can be rewritten in the different manifestly covariant forms \cite{Kr-Sim-1,Krivsky-Sim-2, Broglie,Poland} (see also the review in \cite{Simulik-Nova}).

The 4-vector $\mathcal{E}\equiv (\mathcal{E}^{\mu})$ (and corresponding manifestly covariant Maxwell-like equations) for the first time has been put into consideration in our articles \cite{Mexico,Mexico-2}. Therefore, there is some sense of our suggestion as follows. In addition to the name of Riemann--Silberstein vector for the 3-vector $\overrightarrow{\mathcal{E}}= \overrightarrow{E} -i\overrightarrow{H}$ let us introduce the name Riemann--Silberstein-- Darwin vector for a much more useful 4-vector (\ref{Lab15}). Indeed, C.G. Darwin \cite{Darwin} introduced two 4-component substitutions (see formulae (2.3) in \cite{Darwin} and relations given just below these formulae (2.3)), which transforms the massless Dirac equation into the Maxwell-like system (\ref{General}). We have found \cite{Simulik-1, Simulik-Natur, Mexico} the next six 4-component transitions of that kind. The essence of our step in \cite{Mexico,Mexico-2} is in transition from Darwin's type of substitution $\psi=\mathrm{column}\left|E^{3}+iH^{0}, E^{1}+iE^{2}, iH^{3}+E^{0}, iH^{1}-H^{2}\right|$ to the 4-vector (\ref{Lab15}). Such transition essentially transforms the $\gamma$ matrices in the Dirac equation.

Indeed, in the massless Dirac equation in the terms of (\ref{Lab15})   
\begin{equation}
\label{Dirac}
\widetilde{\gamma}^{\mu}\partial_{\mu}\mathcal{E}=0, \quad \mathcal{E} \equiv E- iH = \left|
{{\begin{array}{*{20}c}
 \overrightarrow{E} -i\overrightarrow{H} \hfill  \\
 E^{0}-iH^{0} \hfill  \\
 \end{array} }} \right|,
\end{equation}
\noindent matrices $\gamma$ belong to the specific Clifford--Dirac algebra $\gamma$-matrix representation
\begin{equation}
\label{Gamma} 
\widetilde{\gamma }^0=\left|
\begin{array}{cccc}
1 & 0 & 0 & 0 \\
0 & 1 & 0 & 0 \\
0 & 0 & 1 & 0 \\
0 & 0 & 0 & -1
\end{array}
\right| C,\quad \widetilde{\gamma }^1=\left|
\begin{array}{cccc}
0 & 0 & 0 & 1 \\
0 & 0 & -i & 0 \\
0 & i & 0 & 0 \\
-1 & 0 & 0 & 0
\end{array}
\right| C,
\end{equation}
$$\widetilde{\gamma }^2=\left|
\begin{array}{cccc}
0 & 0 & i & 0 \\
0 & 0 & 0 & 1 \\
-i & 0 & 0 & 0 \\
0 & -1 & 0 & 0
\end{array}
\right| C,\quad \widetilde{\gamma }^3=\left|
\begin{array}{cccc}
0 & -i & 0 & 0 \\
i & 0 & 0 & 0 \\
0 & 0 & 0 & 1 \\
0 & 0 & -1 & 0
\end{array}
\right| C,$$
\noindent where $C$ is the operator of complex conjugation, $C\mathcal{E}\equiv \mathcal{E}^{*}$ (the operator of involution in the rigged Hilbert space $\mathrm{S}^{3,4}\subset\mathrm{H}^{3,4}\subset\mathrm{S}^{3,4*}$), and the anti-commutation relations $\widetilde{\gamma}^\mu \widetilde{\gamma }^\nu +\widetilde{\gamma }^\nu \widetilde{\gamma }^\mu =2g^{\mu \nu }$ are valid.

Note that systematical investigations of such kind representations of Clifford--Dirac algebra lead us \cite{Simulik-3,Mexico-3} to the discovery of 64-dimensional $\textit{C}\ell^{\mathbb{R}}$(4,2) and $\textit{C}\ell^{\mathbb{R}}$(0,6) $\gamma$-matrix representations of the Clifford algebra over the field of real numbers. On the basis of representation $\textit{C}\ell^{\mathbb{R}}$(0,6) we were able to prove \cite{PhysLett,Lviv-2} the bosonic characteristics of the Dirac equation with nonzero mass, so-called the Fermi--Bose duality of the Dirac equation (the review is given in \cite{Simulik-Nova}). Moreover, in \cite{Simulik-Nova}) the 4-vector (\ref{Lab15}) was used for the formulation of Lagrange approach for the interacting electromagnetic-scalar $\mathcal{E} = E- iH$ and spinor $\Psi$ fields, which is the key for the construction of quantum electrodynamics in the terms of field strengths \cite{Lagrange,Fushchych,KIev}. Listed results demonstrate only small part of the 4-vector $\mathcal{E} = E- iH$ application.

Below we used the form
\begin{equation}
\label{Lab18}
Q_{\mu\nu}\equiv \partial_{\mu}\mathcal{E}_{\nu}-\partial_{\nu}\mathcal{E}_{\mu}+i\varepsilon_{\mu\nu\rho\sigma}\partial^{\rho}\mathcal{E}^{\sigma}=0, \quad \partial_{\mu}\mathcal{E}^{\mu}=0,
\end{equation}
\noindent from \cite{Mexico} and \cite{Kr-Sim-1,Krivsky-Sim-2} for the further study of longitudinal electromagnetic waves. Equations (\ref{Lab18}) can be rewritten as follows

\begin{equation}
\label{Lab19}
\widehat{M}\mathcal{E}\equiv \left|
\begin{array}{cccc}
\partial_{0} & i\partial_{3} & -i\partial_{2} & \partial_{1} \\
-i\partial_{3} & \partial_{0} & i\partial_{1} & \partial_{2} \\
i\partial_{2} & -i\partial_{1} & \partial_{0} & \partial_{3} \\
-\partial_{1} & -\partial_{2} & -\partial_{3} & -\partial_{0}
\end{array}
\right| \left|
\begin{array}{cccc}
\mathcal{E}^{1} \\
\mathcal{E}^{2} \\
\mathcal{E}^{3} \\
\mathcal{E}^{0}
\end{array}
\right|=0.
\end{equation}

The tensor $Q_{\mu\nu}$ in (\ref{Lab18}) has only 3 independent components (it follows from the anti-symmetry of $Q_{\mu\nu}$ and from the Levi-Civita tensor $\varepsilon^{\mu\nu\rho\sigma}$). Therefore, the system (\ref{Lab18}) is the system of four independent equations coinciding with the equations (\ref{Complex-1}) in the special case (\ref{Lab14}), (\ref{Lab15}). The equation (\ref{Lab19}) is the matrix form of the system (\ref{Lab18}). The forms (\ref{Lab18}), (\ref{Lab19}) are useful for the investigations of the symmetry properties of these equations, see our articles \cite{Simulik-1, Broglie, Poland, Mexico} and the book \cite{KIev}.

It follows from (\ref{Lab18}) or (\ref{Lab19}) that every component $\mathcal{E}^{\mu}$ of 4-component function (\ref{Lab15}) obeys the d'Alambert equation $\partial^{\mu}\partial_{\mu}\mathcal{E}^{\nu}=0$. It means in particular that the gradient-like current density (\ref{Lab14}) obeys the continuity equation $\partial_{\mu}j^{\mu}=0$. And in the case of real function $\varphi$ (i.e. if $H^{0}=0$) the current density $j^{\mu}$ is real as well. If in general case $\varphi$ is a complex function (i.e. if $H^{0}\neq 0$) than we have

\begin{equation}
\label{Lab20}
\mathrm{Re}j= (\rho_{\mathrm{E}}, \overrightarrow{j}_{\mathrm{E}}), \quad \mathrm{Im}j= (\rho_{\mathrm{M}}, \overrightarrow{j}_{\mathrm{M}}),
\end{equation}
\noindent where the notations $(\rho_{\mathrm{E}}, \overrightarrow{j}_{\mathrm{E}})$ are used for the ordinary electric charge and current densities. Hypothetical densities of magnetic charge and current are given by $(\rho_{\mathrm{M}}, \overrightarrow{j}_{\mathrm{M}})$.

Below, following \cite{Kr-Sim-1,Krivsky-Sim-2}, we consider the most general case, when $\mathrm{Im}\varphi\neq 0$. Note that, nevertheless, at every step one can put  
\begin{equation}
\label{Lab21}
\mathrm{Im}\varphi=0 \Rightarrow \mathrm{Im}j=0 \Rightarrow \rho_{\mathrm{M}}= \overrightarrow{j}_{\mathrm{M}}=0.
\end{equation}
\noindent Therefore, we have here as a partial case the result of the Section 2 as well.

It is worthwhile to emphasize that even in the case (\ref{Lab21}), the real scalar function $\varphi$ generates electric currents $\overrightarrow{j}_{\mathrm{E}}$ and charges $\rho_{\mathrm{E}}$, which themselves cause the appearance of the electromagnetic field. And the function $\varphi=\mathcal{E}^{0}$ in the equations (\ref{Lab19}) (real or complex) can be either an arbitrarily fixed one (but obeying the equation $\partial^{\mu}\partial_{\mu}\varphi=0$) or even a field variable called below as a scalar (in general, complex) field. 

The assumption that equations (\ref{Lab18})=(\ref{Lab19}) are defined in the rigged Hilbert space $\mathrm{S}^{3,4}\subset\mathrm{H}^{3,4}\subset\mathrm{S}^{3,4*}$ enables one to solve the equations (\ref{Lab19}) by the Fourier method. Therefore, one can find the general solution of these equations with an arbitrary gradient-like current by the straightforward calculations, without appealing to the electromagnetic potentials.

Thus, for the Fourier transform $\widetilde{\mathcal{E}}$ the following equation is valid
\begin{equation}
\label{Lab22}
M(k)\widetilde{\mathcal{E}}=0, \quad k\in \mathrm{R}^{4}(k), \quad \widetilde{\mathcal{E}}\equiv \widetilde{\mathcal{E}}^{\mu}.
\end{equation}
\noindent which is a homogeneous system of four equations for the four functions of the rigged Hilbert space $\mathrm{S}^{3,4}\subset\mathrm{H}^{3,4}\subset\mathrm{S}^{3,4*}$. The calculation of the corresponding determinant leads to the result:  
\begin{equation}
\label{Lab23}
\mathrm{det}M(k) \equiv \mathrm{det}\left|
\begin{array}{cccc}
k_{0} & k_{1} & k_{2} & k_{3} \\
k_{1} & k_{0} & ik_{3} & -ik_{2} \\
k_{2} & -ik_{3} & k_{0} & ik_{1} \\
k_{3} & ik_{2} & -ik_{1} & k_{0}
\end{array}
\right| \equiv (k^{2}_{0} - \vec{k}^{2}).
\end{equation}

Finally, one can found that equations (\ref{Lab19}) have 4 linearly independent partial solutions, and the general solution is the superposition of these partial solutions. The suitable choice of such solutions leads to the expression
\begin{small}
\begin{equation}
\label{Lab24}
\mathcal{E}\left( x\right) =\frac{1}{(2\pi)^{\frac{3}{2}}}\int \mathrm{d}^3k\sqrt{\frac{\omega}{2}}\left\{
\left[ c^{1}_{\vec{k}}e_1+c^{3}_{\vec{k}}\left( e_3+e_4\right) \right] \mathrm{e}^{-ikx}+ \left[ c^{*2}_{\vec{k}}e_1+c^{*4}_{\vec{k}}\left( e_3+e_4\right) \right] \mathrm{e}^{ikx}
\right\} ,
\end{equation}
\end{small}
\noindent where  the functions $\left\{c^{1}_{\vec{k}}, \, c^{2}_{\vec{k}}, \, c^{3}_{\vec{k}}, \, c^{4}_{\vec{k}}\right\}$ of 3-momentum $\vec{k}$ in the corresponding rigged Hilbert space are the complex quantum-mechanical momentum-helicity amplitudes of a photon and spinless massless boson (the amplitudes of the transverse and longitudinal electromagnetic-scalar waves). Further, the notations (\ref{Not5}) are used and the four $4$-component basis vectors $\left\{e\right\}$ are taken in the form
\begin{equation}
\label{Lab25}
e_1=\left|
\begin{array}{c}
\vec{e}_{1} \\
0
\end{array}
\right| ,\quad e_2=\left|
\begin{array}{c}
\vec{e}_{2} \\
0
\end{array}
\right| ,\quad e_3=\left|
\begin{array}{c}
\vec{e}_{3} \\
0
\end{array}
\right| ,\quad e_4=\left|
\begin{array}{c}
0 \\
1
\end{array}
\right|.
\end{equation}
\noindent Here the $3$-component basis vectors $(\vec{e}_{1}, \, \vec{e}_{2}, \, \vec{e}_{3})$ without any loss of generality can be taken as (\ref{Not6}). These objects are the eigen vectors of the quantum-mechanical helicity operator for the spin s = 1.
\begin{equation}
\label{Lab26}
\hat{h}\vec{e}_{h}=h\vec{e}_{h}, \quad \hat{h}\equiv \frac{\vec{s}\cdot \vec{k}}{\omega}, \quad h = (-1,+1,0)\equiv (1,2,3), 
\end{equation}
\noindent where $\vec{s}\equiv (s^j)$ are the generators of irreducible representation $D(1)$ of the group $SU(2)$:
\begin{equation}
s^1=\left|
\begin{array}{ccc}
0 & 0 & 0 \\
0 & 0 & -i \\
0 & i & 0
\end{array}
\right|, \, s^2=\left|
\begin{array}{ccc}
0 & 0 & i \\
0 & 0 & 0 \\
-i & 0 & 0
\end{array}
\right|, \, s^3=\left|
\begin{array}{ccc}
0 & -i & 0 \\
i & 0 & 0 \\
0 & 0 & 0
\end{array}
\right|; \, \vec{s}^2=1(1+1)\mathrm{I_{3}}.
\label{Lab27}
\end{equation}

The formula (\ref{Lab24}) represents the exact solution of the Maxwell equations with specific kind of inhomogeneity determined by the gradient-like sources (the unusual magnetic sources can be easily put equal to zero, in this case we have $c^{3}_{\vec{k}}=c^{4}_{\vec{k}}$). The presentation of the general solution (\ref{Lab24}) separately for $\overrightarrow{\mathcal{E}}= \overrightarrow{E} -i\overrightarrow{H}$ and $\mathcal{E}^{0}=E^{0}-iH^{0}\equiv\varphi$ ensures some visualization of the result:
\begin{small}
\begin{equation}
\label{Lab28}
\overrightarrow{\mathcal{E}}\left(x\right) =\frac{1}{(2\pi)^{\frac{3}{2}}}\int\mathrm{d}^3k\sqrt{\frac{\omega}{2}}\left[c^{1}_{\vec{k}}\mathrm{e}^{-ikx}+c^{*2}_{\vec{k}}\mathrm{e}^{ikx}\right]\vec{e}_{1}+ \left[c^{3}_{\vec{k}}\mathrm{e}^{-ikx}+c^{*4}_{\vec{k}}\mathrm{e}^{ikx}\right]\vec{e}_{3},
\end{equation}
\end{small}
\begin{equation}
\label{Lab29}
\varphi \equiv E^0-iH^0=\frac{1}{(2\pi)^{\frac{3}{2}}}\int \mathrm{d}^3k\sqrt{\frac{\omega}{2}}\left(c^{3}_{\vec{k}}\mathrm{e}^{-ikx}+c^{*4}_{\vec{k}}\mathrm{e}^{ikx}\right).
\end{equation}

Hence, the general solution of the Maxwell equations (\ref{Complex-1}) with an arbitrary gradient-like current (\ref{Lab14}) contains both transverse and longitudinal parts
\begin{equation}
\label{Lab30}
\overrightarrow{\mathcal{E}}\left(x\right)=\overrightarrow{\mathcal{E}}^{\mathrm{tr}}\left(x\right)+\overrightarrow{\mathcal{E}}^{\mathrm{lon}}\left(x\right).
\end{equation}
\noindent Of course, in the case
\begin{equation}
\label{Lab31}
\varphi=0 \Rightarrow \rho=\overrightarrow{j}=0 \quad c^{3}_{\vec{k}}=c^{*4}_{\vec{k}}=0,
\end{equation}
\noindent the solution of the Maxwell equations (\ref{Complex-1}) contains only a superposition (\ref{Sol2}) of the transverse waves $\vec{e}_{1}=\mathrm{exp}(\mp)ikx$ (left-hand circularly polarized photons, which are a basis for the irreducible D(0,1) representation of the Lorentz group SL(2,C)). Non-vanishing gradient-like currents cause only the appearance of longitudinal waves $\vec{k}/\omega =\mathrm{exp}(\mp)ikx$. Moreover, if the scalar field $\varphi$ (and, consequently, the gradient-like current), related to the electromagnetic field $\overrightarrow{\mathcal{E}}= \overrightarrow{E} -i\overrightarrow{H}$ according to the Maxwell equations (\ref{Lab18}), (\ref{Lab19}), is nonzero in an asymptotically big space-time region, than the longitudinal waves propagate in the same asymptotically big space-time region. Indeed, the gradient-like current densities (\ref{Lab14}) and the solutions (\ref{Lab28}), (\ref{Lab29}) are determined by the one and the same amplitudes.

It should be marked that all the assertions of this paper arc related to the case of space, where the relation $\varepsilon=\mu =1$ for the dielectric constant $\varepsilon$ and magnetic permeability $\mu$ is valid. In the case of nontrivial functions $\varepsilon$ and $\mu$, i. e. in the case of inhomogeneous medium, the interaction between $\varphi$ and $\overrightarrow{\mathcal{E}}= \overrightarrow{E} -i\overrightarrow{H}$ may give the nonlinear effects. Indeed, in the inner-atomic medium considered in the book \cite{Simulik-Montreal} and in original articles \cite{Mexico-2,Simulik-UkrMath} the scalar field $\varphi \equiv \mathcal{E}^0 =E^0-iH^0$ interacting with the electromagnetic field $\overrightarrow{\mathcal{E}}= \overrightarrow{E} -i\overrightarrow{H}$ (via the gradient-like current (\ref{Lab14})) is appeared to be so closely related to the field $\overrightarrow{\mathcal{E}}= \overrightarrow{E} -i\overrightarrow{H}$ that from $\varphi=0$ it follows $\overrightarrow{\mathcal{E}}= \overrightarrow{E} -i\overrightarrow{H}=0$ as well. Such case is considered by H. Sallhofer \cite{Sall-2,Sall-3} and in our papers \cite{Broglie,Poland} dealing with the closed coupled electromagnetic-scalar waves.

An interesting, but a "draft", hypothesis on longitudinal electric waves is suggested recently in \cite{Spirichev}). It is shown that the longitudinal waves of the divergence of the vector potential propagate at a speed greater than the speed of light and do not have a magnetic component.

Let us mention that the electrodynamics with magnetic currents and charges considered in this section is some natural generalization of conventional
Maxwell's electrodynamics, i. e., in general, we considered unusual electrodynamics here. Nevertheless the existence of longitudinal electromagnetic
waves does not depend on the presence of magnetic currents and charges. Let us emphasize that in the case (\ref{Lab21}), when the magnetic sources are put equal to zero and, therefore, $c^{3}_{\vec{k}}=c^{4}_{\vec{k}}$, the longitudinal electromagnetic waves still exist. We have in this case $\overrightarrow{H}^{\mathrm{lon}}=0$ but $\overrightarrow{E}^{\mathrm{lon}}\neq 0$. This fact is evident after rewriting the solutions (\ref{Lab28}), (\ref{Lab29}) in the terms of real electromagnetic field strengths ($\overrightarrow{E}, \overrightarrow{H}$), see below. Therefore, the longitudinal electric waves can exist within the framework of ordinary Maxwell's electrodynamics (without magnetic monopoles) in the case, when the electric currents and charges have specific gradient-like form.

Note that if the quantities $E^0,H^0$ in equations (\ref{General}) are some given functions, for which the representation (\ref{Lab29}) is valid, then (\ref{General}) are the Maxwell equations with the given sources, $j_\mu ^{\mathrm{E}}=-\partial _\mu E^0, \, j_\mu ^{\mathrm{M}}=-\partial _\mu H^0$ (namely these 4 currents we call the gradient-like sources). In this case the general solution of the Maxwell equations (\ref{General}), (\ref{Lab18}), (\ref{Lab19}) with the given sources, as follows from (\ref{Lab28}), (\ref{Lab29}), has the form

\begin{small}
$$\overrightarrow{E}(x)=\frac{1}{(2\pi)^{\frac{3}{2}}}\int d^{3}k\sqrt{\frac{\omega}{2}}\left\{\left[c^{1}_{\vec{k}}\vec{e}_{1}+c^{2}_{\vec{k}}\vec{e}_{2}+(c^{3}_{\vec{k}}+c^{4}_{\vec{k}})\vec{e}_{3}\right]e^{-ikx}+A\right\},$$
\end{small}
$$A \equiv \left[c^{*1}_{\vec{k}}\vec{e}_{1}^{\,*}+c^{*2}_{\vec{k}}\vec{e}_{2}^{\,*}+(c^{*3}_{\vec{k}}+c^{*4}_{\vec{k}})\vec{e}_{3}^{\,*}\right]e^{ikx},$$
\begin{small}
\begin{equation}
\label{Lab32}\overrightarrow{H}(x)= \frac{i}{(2\pi)^{\frac{3}{2}}}\int d^{3}k\sqrt{\frac{\omega}{2}}\left\{\left[c^{1}_{\vec{k}}\vec{e}_{1} -c^{2}_{\vec{k}}\vec{e}_{2}+(c^{3}_{\vec{k}}-c^{4}_{\vec{k}})\vec{e}_{3}\right]e^{-ikx}-B\right\},
\end{equation}
\end{small}
$$B \equiv \left[c^{*1}_{\vec{k}}\vec{e}_{1}^{\,*}-c^{*2}_{\vec{k}}\vec{e}_{2}^{\,*}+(c^{*3}_{\vec{k}}-c^{*4}_{\vec{k}})\vec{e}_{3}^{\,*}\right]e^{ikx},$$
$$E^{0}(x)=\frac{1}{(2\pi)^{\frac{3}{2}}}\int d^{3}k\sqrt{\frac{\omega}{2}}\left[(c^{3}_{\vec{k}}+c^{4}_{\vec{k}})e^{-ikx}+(c^{*3}_{\vec{k}}+c^{*4}_{\vec{k}})e^{ikx}\right],$$
$$H^{0}(x)=\frac{i}{(2\pi)^{\frac{3}{2}}}\int d^{3}k\sqrt{\frac{\omega}{2}}\left[(c^{3}_{\vec{k}}-c^{4}_{\vec{k}})e^{-ikx}-(c^{*3}_{\vec{k}}-c^{*4}_{\vec{k}})e^{ikx}\right].$$
\noindent Here both the electric field strength $\overrightarrow{E}(x)$ and the magnetic field strength $\overrightarrow{H}(x)$ contain (together with ordinary transverse waves) the corresponding longitudinal waves as well. These longitudinal electric and longitudinal magnetic waves are given by the amplitudes $c^{3}_{\vec{k}}+c^{4}_{\vec{k}}$ and $c^{3}_{\vec{k}}-c^{4}_{\vec{k}}$, respectively. The scalar functions $(E^{0}(x), \, H^{0}(x))$ specify the electromagnetic currents and charges densities in the Maxwell-like system of equations (\ref{General}). The scalar waves $(E^{0}(x), \, H^{0}(x))$ propagate in the direction of $\vec{e}_{3}=\vec{k}/\omega$ as well.

The conservation laws of energy and momentum of this system are given by
\begin{equation}
\label{Lab33}
P^0=\frac 12\int d^3x\mathcal{E}^{\dagger }\mathcal{E}=\frac 12\int d^3x(\overrightarrow{E}^2+\overrightarrow{H}^2+E_0^2+H_0^2),
\end{equation}
$$\overrightarrow{P}_{gen}=\int d^3x(\overrightarrow{E}\times \overrightarrow{H}-\overrightarrow{E}E^0-\overrightarrow{H}H^0),$$
\noindent where the useful notation $\mathcal{E}$ is explained below in the formula (\ref{Lab15}). The normalization factor in (\ref{Lab32}) is taken from the demand that the conserved quantities in (\ref{Lab33}) are taken in the dimension of energy and momentum of the system of electromagnetic and scalar fields.

In our articles \cite{Broglie, Poland}, the system of equations (\ref{General}) is called as the slightly generalized Maxwell equations with gradient-type sources. Indeed, at the first step, the system (\ref{General}) is the generalization of the Maxwell equations because it contains the condition $\mathrm{div}\overrightarrow{H}\neq 0$ and the nonzero magnetic current density in the equation $\partial_{0}\overrightarrow{H}+\mathrm{curl}\overrightarrow{E}=-\mathrm{grad}H^{0}$. Nevertheless, at the second step the system (\ref{General}) is the simplification (specification) of such Maxwell equations. Corresponding electromagnetic currents and charge densities are the partial gradient-type forms of the general form of the electromagnetic sources.

Thus, the system of equations (\ref{General}) due to the conditions $\partial_{0}\overrightarrow{H}+\mathrm{curl}\overrightarrow{E}\neq 0, \, \mathrm{div}\overrightarrow{H}\neq 0$ is not the standard Maxwell electrodynamics. Therefore, it is better to start the experimental testing of the longitudinal electromagnetic waves by the experimental modeling of the situation given by the system (\ref{Main}), which belongs to the standard Maxwell electrodynamics. This system of equations predicts an existence of the longitudinal component of the vector of electric field strength $\overrightarrow{E}$. 

The arguments in the prospect of system (\ref{General}) are not so evident. Nevertheless, let us mention (i) that system (\ref{General}) has the maximally possible symmetry properties among the Maxwell and the Maxwell-like systems of equations. Indeed, in \cite{Broglie, Poland} the 256 dimensional algebra of invariance of equations (\ref{General}) was mentioned. And the equations (\ref{General}) have been considered in the terms of complex function (\ref{Lab15}). Taking into account the symmetries found recently in \cite{PhysLett, Lviv-2}, the number 256 can be increased. Let us recall that the role of the symmetry principle in electrodynamics is known from the times of Maxwell and Heaviside.

(ii) The system of equation (\ref{General}) has the property of the Fermi--Bose duality. It can describe (\cite{Broglie, Poland}) both the massless spin 1/2 particle-antiparticle doublet of the fermions and the massless spin (1,0) doublet of bosons (photon and massless spinless boson). In the terms of complex 4-vector (\ref{Lab15}) equations (\ref{General}) has the manifestly covariant form (\ref{Lab18}) together with the form of massless Dirac equation (\ref{Dirac}) with Clifford--Dirac algebra $\gamma$ matrix representation (\ref{Gamma}) described in \cite{Mexico-3}.

(iii) The system of equations (\ref{General}) considered in the specific inhomogeneous medium \cite{Sall-2, Sall-3} can be applied to inneratomic phenomena and describe the hydrogen spectrum (another approach to the hydrogen spectrum description on the basis of stationary Maxwell equations in this medium is given in \cite{Broglie, Poland, Mexico-2, Simulik-Montreal, Simulik-UkrMath}). In \cite{Sall-2, Sall-3} and \cite{Simulik-Montreal, Simulik-UkrMath} the role of longitudinal components is evident. Note that in the inhomogeneous medium \cite{Broglie, Poland, Mexico-2, Simulik-Montreal, Simulik-UkrMath} the solutions of the Maxwell equations (\ref{General}) are given by the closed in sphere beams of the electromagnetic waves. 

Therefore, there is some sense in experimental modeling of the situation given by the system (\ref{General}) in the problem of the longitudinal electromagnetic waves investigation. Of course, the advantages of system (\ref{Main}), which belongs to the standard electrodynamics, are evident.

Note that solution (\ref{Sol-1}) can be found both by the direct application of the Fourier method and as a partial case of solution (\ref{Lab32}).

The validity of solution (\ref{Lab32}) can be verified by the direct substitution of (\ref{Lab32}) into the equations (\ref{General}).

\section{Comments on the paper of K.J. van Vlaenderen and A. Waser}\label{Waser}

The paper of K.J. van Vlaenderen and A. Waser \cite{Hadronic} contains the main result of our articles \cite{Kr-Sim-1,Krivsky-Sim-2} without any reference on it. This result is formulated in \cite{Hadronic} as "Generalization of classical electrodynamics to admit a scalar field and longitudinal waves". Nevertheless, this result has been published already a few years before in \cite{Kr-Sim-1,Krivsky-Sim-2}. Moreover, in \cite{Kr-Sim-1,Krivsky-Sim-2} we have proved also the existence of longitudinal electric and scalar waves in the framework of standard classical electrodynamics (see also \cite{Simulik-2, Simulik-Gor-Z} for further information). Therefore, "the generalization of classical electrodynamics" suggested in \cite{Hadronic} is not necessary to admit a longitudinal electric and scalar waves. 

During the middle part of August 2017 two International conferences were held in Uzhgorod in the Institute of Electron Physics of NAS of Ukraine: "Non-Euclidean geometry in modern physics, BGL-1", Uzhgorod, Ukraine, 13-16 of August 1997, and "The centenary of electron", EL-100, Uzhgorod, Ukraine, 18-20 of August 1997. These two conferences were combined into one and held together. Our report "General solution of the Maxwell equations with gradient-like sources and longitudinal electromagnetic waves" \cite{Kr-Sim-1} aroused the interest of many participants from different countries. The conference on non-Euclidean geometry in modern physics in well-known today. In 2019 was held the XI conference of this series.

Moreover, the result under consideration was the part of my thesis for a full doctor's degree "Relationships between spin 1 and 1/2 fields and their role in construction of quantum electrodynamics in the terms of field strengths", which defense took place on 25 of April 2000 at the Kiev National University. The defense passed a wide public discussion of the dissertation at the National Academy of Sciences of Ukraine (Bogolyubov Institute for Theoretical Physics, Kiev, Institute of Physics, Kiev, Institute for Condensed Matter Physics, Lviv, Institute of Electron Physics, Uzhgorod).  

Thus, our assertion about \cite{Hadronic} is as follows. We cannot speak about evident plagiarism, but we can mark exactly that the main results of \cite{Hadronic} were found by us before in \cite{Kr-Sim-1,Krivsky-Sim-2} and were well known. Moreover, the interpretation in \cite{Kr-Sim-1,Krivsky-Sim-2} is more adequate: "the longitudinal electro-scalar wave can be described in the framework of standard classical electrodynamics without any generalization". As long as \cite{Hadronic} has today the followers we pay attention of our colleagues for the conclusions in \cite{Kr-Sim-1,Krivsky-Sim-2} and here. The "wave of information" coming into literature from \cite{Hadronic} on the generalization of Maxwell electrodynamics is not necessary. The standard classical electrodynamics is working well in taking into account longitudinal electric and corresponding scalar waves. We do not list the names of adherents of \cite{Hadronic} that were misled by inaccuracies in this article. We have a hope that consideration in \cite{Kr-Sim-1,Krivsky-Sim-2} and here will be useful for them as well.

In \cite{Simulik-Gor-Z}, and in the report at the corresponding international conference, we already mentioned briefly that main results of \cite{Hadronic} were known from \cite{Kr-Sim-1,Krivsky-Sim-2}. Finally, the PDF files of publications \cite{Kr-Sim-1,Krivsky-Sim-2} can be found during many years on the Internet, e.g., it is enough to click the title "General solution of the Maxwell equations with gradient-like sources and longitudinal electromagnetic waves".

\section{The subsystems of the Maxwell-like equations with gradient-type electric and magnetic sources}\label{subsystems}

Due to the above mentioned properties, the system of equations (\ref{Main}) is the most interesting subsystem of the Maxwell-like equations (\ref{General}) (in the problem of longitudinal electromagnetic waves consideration). Equations (\ref{Main}) follow from equations (\ref{General}) after substitution $H^{0}=0 \, \Rightarrow c^{3}_{\vec{k}} = c^{4}_{\vec{k}}$. In this case, in solution (\ref{Sol-1}) the amplitude of the longitudinal wave $\alpha_{\vec{k}}= 2c^{3}_{\vec{k}}$. This partial case of (\ref{General}) is presented in section 2 in details.

Another interesting subsystem of (\ref{General}) follows from (\ref{General}) after the substitution $E^{0}(x)=0$. In this case, the vector $\overrightarrow{H}(x)$ of the magnetic field strength contains the longitudinal wave component. Other properties of the general solution are similar to (\ref{Sol-1}) and are to (\ref{Sol-1}) in evident symmetry. Nevertheless, the interest to this case is lower then for $H^{0}=0$ case. The conditions $\mathrm{div}\overrightarrow{H}\neq 0$ and $\partial_{0}\overrightarrow{H}+\mathrm{curl}\overrightarrow{E}\neq 0$ are not in the framework of the standard Maxwell electrodynamics.

\section{Conclusions and some general comments}\label{conclusions}

\textbf{Purpose}. The long time discussion on existence, or not existence, of longitudinal electromagnetic waves both in nature and in the Maxwell classical electrodynamics is under consideration. The modern experiments on the existence of such waves are reviewed briefly. The link between the longitudinal electromagnetic waves and the system of Maxwell equations is demonstrated.

\textbf{Methods}. Maxwell classical electrodynamics, Fourier method, Fourier transform, amplitude analysis, mapping of the Dirac theory, analysis of the experiments.

\textbf{Main results}. The longitudinal wave component of the electric field strength vector $\overrightarrow{E}(x)$ is found as an exact solution of the standard Maxwell equations with specific current and charge densities (\ref{Sources}) of the gradient type. The corresponded scalar wave component, which is propagated in the same direction, is found as well. Thus, the link between the longitudinal electric waves and the standard Maxwell electrodynamics is demonstrated. In addition, the longitudinal components of both electric and magnetic field strengths, together with two corresponded scalar waves, are found as the exact solution of generalized Maxwell equations, which are characterized by the maximally high symmetry properties.

\textbf{Hypothesis and suggestions}. The calculation of $\rho(x)$ and $\overrightarrow{j}(x)$ according to (\ref{Sources}) and (\ref{Sol-1}) results in oscillating charge density with amplitudes $\alpha_{\vec{k}}, \, \alpha^{*}_{\vec{k}}$ and frequency $\omega$ (together with corresponding 3-current density). We have a hope that it can be useful for the modeling of electrons, positrons, ions, clusters and other charge particles of plasma. Thus, the different processes in plasma (plasmons) in general and longitudinal electromagnetic waves in these objects in particular can be described. Another suggestion is that longitudinal electric wave in \cite{Miyaji,Dorn,Niziev,Tidwell} and in \cite{Kuzn} appears after the interaction of circularly polarized transverse electromagnetic waves.

\textbf{Main Conclusions}. The analysis of found solutions (\ref{Sol-1}) demonstrates that longitudinal components are located near the corresponded current and charge densities, which are the sources of such fields. It follows from the fact that current and charge densities (\ref{Sources}) and the corresponded longitudinal components in the solutions (\ref{Sol-1}) are determined by the same amplitudes $\alpha_{\vec{k}}, \, \alpha^{*}_{\vec{k}}$. The best examples of corresponding physical reality are such big charges as the whole water area of closed sea, the planet Earth in general, their oscillations and corresponding longitudinal electric and scalar waves. The validity of the solution (\ref{Sol-1}) can be verified by the direct substitution of (\ref{Sol-1}) into equations (\ref{Main}). The author has a hope that corresponding experimental situation can be created by modeling the sources (\ref{Sources}) and may be useful. Moreover, all considered here experiments are not in contradiction with such longitudinal electric and scalar waves. The subclass of the functions $\overrightarrow{j}(x), \, \rho(x)$ from (\ref{Sources}) is generated by the oscillating charges in plasma, plasmons, laser beams \cite{Miyaji,Dorn,Niziev,Tidwell}, antennas \cite{Umul}.

Thus, the results are useful for the interpretation of the experiments on the registration of the longitudinal electric waves generating by the interacting laser beams \cite{Miyaji,Dorn,Niziev,Tidwell}, transverse waves in waveguides \cite{Kuzn}, detected by the ball antenna \cite{Monstein}, contributed to longitudinal electromagnetic oscillations in metals \cite{Dats}. Theoretical models of existence of longitudinal electric waves in plasma \cite{Kovr, Bogdan}, plasmons from QED vacuum \cite{Petrov}, crystals, where excitons are produced \cite{Pekar}, radiated antenna \cite{Umul} are the subject of our model application as well.

Finally, all the solutions of equations are found here in the terms of field strengths of electromagnetic (or only electric) and scalar fields without appealing to the potentials. Indeed, in classical electrodynamics the observables are the field strengths itself and only the differences of potentials.

\section*{Acknowledgments}

The author is much grateful to the Prof. N.P. Khvorostenko, who was able many years ago in the far away year 1991 to inspire forever my interest to the problem of longitudinal electromagnetic waves.
I am especially grateful to Prof. D.B. Kuryliak for the useful discussions.





\end{document}